\documentclass[12pt]{article}

\begin{document}

\title{\bf On Physical Properties of Cylindrically Symmetric
Self-Similar Solutions}

\author{M. Sharif \thanks{hasharif@yahoo.com} and
Sehar Aziz \thanks{sehar$\_$aziz@yahoo.com}\\
Department of Mathematics, University of the Punjab,\\
Quaid-e-Azam Campus, Lahore-54590, Pakistan.}

\date{}

\maketitle

\begin{abstract}
This paper is devoted to discuss some of the features of
self-similar solutions of the first kind. We consider the
cylindrically symmetric solutions with different homotheties. We
are interested in evaluating the quantities acceleration,
rotation, expansion, shear, shear invariant and expansion rate.
These kinematical quantities are discussed both in co-moving as
well as in non-co-moving coordinates (only in radial direction).
Finally, we would discuss the singularity feature of these
solutions. It is expected that these properties would help in
exploring some interesting features of the self-similar solutions.
\end{abstract}


\section{Introduction}
Einstein field equations (EFEs) given by
\begin{eqnarray}
R_{ab}-\frac{1}{2}g_{ab}R=\kappa T_{ab},\quad (a,b=0,1,2,3)
\end{eqnarray}
where $R_{ab},~R,~T_{ab},~\kappa$ are the Ricci tensor, Ricci
scalar, matter tensor and the gravitational constant respectively,
are highly non-linear partial differential equations. To simplify
these equations, we frequently impose some symmetry on the
concerned system. Self-similarity is very helpful in simplifying
the field equations. In Newtonian gravity or General Relativity
(GR), there does not exist any characteristic scale. A set of
field equations remains invariant under a scale transformation for
an appropriate matter field. This indicates that there exist scale
invariant solutions to the EFEs. These solutions are known as
self-similar solutions.

We study self-similar solutions because of its two important
features. The first special feature of self-similar solutions is
that, by a suitable coordinate transformations, the number of
independent variables can be reduced by one and hence reduces the
field equations. This variable is a dimensionless combination of
the independent variables, namely the space coordinates and the
time. In other words, self-similarity refers to an invariance
which simply allows the reduction of a system of partial
differential equations to ordinary differential equations.
Secondly, self-similar solutions play an important role in
describing the asymptotic properties of more general models. For
example, the expansion of universe from big bang and the critical
and gravitational collapse of a star might have self-similarity in
some sense as we expect that the initial conditions have been
lost.

In order to obtain realistic solutions of gravitational collapse
leading to star formation, self-similar solutions have been
investigated by many authors in Newtonian gravity [1]. However, in
GR, these solutions were first studied by Cahill and Taub [2].
They studied these solutions in the cosmological context and under
the assumption of spherically symmetric distribution of a
self-gravitating perfect fluid. In GR, self-similarity is defined
by the existence of a homothetic vector (HV) field. Such
similarity is called the first kind (or homothety or continuous
self-similarity). There exists a natural generalization of
homothety called kinematic self-similarity, which is defined by
the existence of a kinematic self-similar (KSS) vector field. A
KSS vector $\xi$ satisfies the following conditions
\begin{eqnarray}
\pounds_{\xi}h_{ab}&=& 2\delta h_{ab},\\
\pounds_{\xi}u_{a}&=& \alpha u_{a},
\end{eqnarray}
where $h_{ab}=g_{ab}+u_au_b$ is the projection tensor, $\alpha$
and $\delta$ are constants. The similarity transformation is
characterized by the scale independent ratio, $\alpha/\delta$,
which is known as the similarity index and can be classified into
three kinds.

The idea of self-similarity by Cahil and Taub corresponded to
Newtonian self-similarity of the homothetic class. Carter and
Henriksen [3,4] defined the other kinds of self-similarity namely
second, zeroth and infinite kind. In the context of kinematic
self-similarity, homothety is considered as the first kind.
Several authors have explored KSS perfect fluid solutions. The
only barotropic equation of state which is compatible with
self-similarity of the first kind is $p=k\rho$.

Carr [5] has classified the self-similar perfect fluid solutions
of the first kind for the dust case ($k=0$). The case $0<k<1$ has
been studied by Carr and Coley [6]. Coley [7] has shown that the
FRW solution is the only spherically symmetric homothetic perfect
fluid solution in the parallel case. McIntosh [8] has discussed
that a stiff fluid ($k=1$) is the only compatible perfect fluid
with the homothety in the orthogonal case. Benoit and Coley [9]
have studied analytic spherically symmetric solutions of the EFEs
coupled with a perfect fluid and admitting a KSS vector of the
first, second and zeroth kinds.

Carr et al. [10] have considered the KSS associated with the
critical behavior observed in the gravitational collapse of
spherically symmetric perfect fluid with equation of state
$p=k\rho$.  Carr et al. [11], further, investigated solution space
of self-similar spherically symmetric perfect fluid models and
physical aspects of these solutions. They combine the state space
description of the homothetic approach with the use of the
physically interesting quantities arising in the co-moving
approach. Coley and Goliath [12] have investigated self-similar
spherically symmetric cosmological models with a perfect fluid and
a scalar field with an exponential potential.

Recently, Maeda et al. [13,14] investigated the KSS vector of the
second kind in the tilted case. In their recent paper [15], the
same authors discussed the classification of the spherically
symmetric KSS perfect fluid and dust solutions. The existence of
self-similar solutions of the first kind is related to
conservation laws and to the invariance of the problem with
respect to the group of similarity transformations of quantities
with independent dimensions. This can be characterized in GR by
the existence of a homothetic vector.

Qadir et al. [16] have classified cylindrically symmetric static
manifold according to their homotheties and metrics. In each case
the homothety vector field and the corresponding metrics are
obtained explicitly by solving the homothety equations. They found
10 solutions admitting 4, 5, 7 or 11 homotheties. There is only
one metric admitting 7 homotheties and three metrics correspond to
4, three correspond to 5 and three correspond to 11 homotheties.

In a recent paper, Sharif and Sehar [17] have explored some
physical properties of the spherically symmetric self-similar
solutions of the first kind. This has provided some interesting
features of self-similar solutions. In this paper, we are
extending the same analysis for the cylindrically symmetric
self-similar solutions of the first kind. The paper can be
outlined as follows. In section 2, we shall write down the
self-similar solutions of cylindrically symmetric spacetimes.
Section 3 is devoted for the discussion of the physical properties
of these self-similar solutions of the first kind both in
co-moving and non-co-moving coordinates. In section 4, we shall
explore the singularity feature of these solutions. Finally, we
shall summarize and discuss all the results.

\section{Cylindrically Symmetric self-similar solution of first kind}

The general cylindrically symmetric static spacetime is given by
the line element [18]
\begin{equation}
ds^2=e^{\nu(r)}dt^2- dr^2-a^2 e^{\lambda(r)}d\theta^2-
e^{\mu(r)}dz^2,
\end{equation}
where $\nu$, $\lambda$ and $\mu$ are arbitrary functions of $r$
and $a$ has units of length.

Qadir et al. [17] have solved the homothetic equations and found
self-similar solutions of the first kind (i.e. homothetic
solutions). There are four classes of such solutions admitting 4,
5, 7 or 11 homotheties respectively. The class $A$ admitting 4
homotheties has three metrics.

The first metric is of Petrov type $I$, Segr\'{e} type [1,111]
[19] and is given by
\begin{equation}
ds^2=e^{2\alpha\ln(\frac{r}{r_0})}dt^2- dr^2-a^2
e^{2\ln(\frac{r}{a})}d\theta^2- dz^2,\quad (\alpha\neq0,1).
\end{equation}
The second metric is of Petrov type $D$, Segr\'{e} type [1,(11)1]
and it follows
\begin{eqnarray}
ds^2=e^{2\alpha\ln(\frac{r}{r_0})}dt^2- dr^2-a^2
e^{2\beta\ln(\frac{r}{r_0})}d\theta^2-e^{2\beta\ln(\frac{r}{r_0})}
dz^2,\nonumber\\
(\alpha\neq\beta, \quad \alpha,\beta\neq0,1).
\end{eqnarray}
The third metric in this class is of Petrov type $I$, Segr\'{e}
type [1,(11)1] and takes the form
\begin{equation}
ds^2=e^{2\alpha\ln(\frac{r}{r_0})}dt^2-dr^2-a^2d\theta^2
-e^{2\ln(\frac{r}{r_0})} dz^2.
\end{equation}
The class $B$ of homothetic solutions corresponding to 5
homotheties also has three metrics. The first and the second
solutions are of Petrov type $D$, Segr\'{e} type [1,111] and are
given by
\begin{eqnarray}
ds^2=e^{2\ln(\frac{r}{r_0})}dt^2- dr^2-a^2
e^{2\ln(\frac{r}{a})}d\theta^2- dz^2,\\
ds^2=e^{2\ln(\frac{r}{r_0})}dt^2- dr^2-a^2
d\theta^2-e^{2\ln(\frac{r}{r_0})}dz^2.
\end{eqnarray}
The third metric is of Petrov type $D$, Segr\'{e} type [(1,1)(11)]
and has the form
\begin{equation}
ds^2=dt^2-dr^2-a^2e^{2\ln(\frac{r}{a})}d\theta^2
-e^{2\ln(\frac{r}{a})} dz^2.
\end{equation}
The class $C$ admitting 7 homotheties corresponds to only one
metric which is of Petrov type $D$, Segr\'{e} type [1,(11)1] and
is given by
\begin{equation}
ds^2=e^{2\alpha\ln(\frac{r}{r_0})}dt^2-dr^2
-a^2e^{2\alpha\ln(\frac{r}{r_0})}d\theta^2-
e^{2\alpha\ln(\frac{r}{r_0})}dz^2.
\end{equation}
This metric represents a tachyonic fluid and could be
re-interpreted as an anisotropic fluid with an appropriately
chosen cosmological constant.

Finally, in the class, $D$ there are three metrics admitting 11
homotheties other than Minkowski space. These are given by the
following three metrics
\begin{eqnarray}
ds^2=dt^2-dr^2 -a^2d\theta^2- e^{2\ln(\frac{r}{r_0})}dz^2,\\
ds^2=dt^2-dr^2 -a^2d\theta^2- dz^2,\\
ds^2=e^{2\ln(\frac{r}{r_0})}dt^2-dr^2 -a^2d\theta^2- dz^2.
\end{eqnarray}

\section{Kinematics of the velocity field}

In this section, we shall discuss some of the kinematical
properties of the self-similar solutions of the first kind, given
by Eqs.(5)-(14), both in co-moving and non-co-moving coordinates.
The kinematical properties [18] to be discussed can be listed as
follows. The acceleration can be defined as
\begin{equation}
\dot{u}_a=u_{a;b} u^b.
\end{equation}
The rotation is given by
\begin{equation}
\omega_{ab}=u_{[a;b]}+ \dot{u}_{[a}u_{b]}.
\end{equation}
The volume behavior of the fluid can be determined by the
expansion scalar defined by
\begin{equation}
\Theta=u^{a}_{;a}.
\end{equation}
The shear tensor, which provides the distortion arising in the
fluid flow leaving the volume invariant, can be found by
\begin{equation}
\sigma_{ab}=u_{(a;b)}+\dot{u}_{(a}u_{b)} -\frac{1}{3}\Theta
h_{ab}.
\end{equation}
Shear scalar, which gives the measure of anisotropy, is defined as
\begin{equation}
\sigma=\sigma_{ab}\sigma^{ab}.
\end{equation}
The rate of change of expansion with respect to proper time is
given by Raychaudhuri's equation [19]
\begin{equation}
\frac{d\Theta}{d\tau}=-\frac{1}{3}\Theta^2-\sigma_{ab}\sigma^{ab}
+\omega_{ab}u^{a}u^{b}-R_{ab}u^{a}u^{b}.
\end{equation}
Now we discuss these properties for the above mentioned
self-similar solutions.

\subsection{Kinematic Properties in Co-moving Coordinates}

First we evaluate the kinematical properties of the self-similar
solutions in co-moving coordinates. We discuss these properties
for all the classes of metrics mentioned above.

\subsubsection{Class $A$ admitting four homotheties}

There are three metrics in this class given by Eqs.(5)-(7). In
co-moving coordinates, the five quantities, i.e., acceleration,
rotation, expansion scalar, shear and shear invariant turn out to
be same for the three metrics in this class. The expansion scalar
is zero for this class. The non-zero component of acceleration is
\begin{equation}
\dot{u}_1=-\frac{\alpha}{r}.
\end{equation}
The non-vanishing rotation component takes the form
\begin{equation}
\omega_{01}=2\frac{\alpha}{r}e^{\alpha\ln(\frac{r}{r_0})}
\end{equation}
and the shear component is
\begin{equation}
\sigma_{01}=-\omega_{01}.
\end{equation}
Finally, the shear invariant becomes
\begin{equation}
\sigma=-4\frac{\alpha^2}{r^2}.
\end{equation}
For this class, the rate of change of expansion is different for
all the three solutions. Using Raychaudhuri's equation, given by
Eq.(20), the rate of expansion for these solutions turns out to be
\begin{eqnarray}
\frac{d\Theta}{d\tau}&=&\frac{3\alpha^2}{r^2},\\
\frac{d\Theta}{d\tau}&=&\frac{\alpha}{r^2}(1-2\beta+3\alpha),\\
\frac{d\Theta}{d\tau}&=&\frac{\alpha}{r^2}(1+3\alpha)
\end{eqnarray}
respectively.

\subsubsection{Class $B$ admitting five homotheties}

For this class, all the kinematic quantities turn out to be the
same for the two metrics given by Eqs.(8)-(9). For these two
solutions, the expansion scalar vanishes. The only component of
acceleration is
\begin{equation}
\dot{u}_1=-\frac{1}{r}
\end{equation}
while the rotation component is given by
\begin{equation}
\omega_{01}=\frac{2}{r}e^{\ln(\frac{r}{r_0})}.
\end{equation}
The non-zero component of shear takes the form
\begin{equation}
\sigma_{01}=-\omega_{01}
\end{equation}
and the shear invariant becomes
\begin{equation}
\sigma=-\frac{4}{r^2}.
\end{equation}
The rate of expansion turns out to be
\begin{equation}
\frac{d\Theta}{d\tau}=\frac{3}{r^2}.
\end{equation}
All the quantities vanish for the third solution given by Eq.(10).

\subsubsection{Class $C$ admitting seven homotheties}

There is only one metric in this class given by Eq.(11) The
expansion scalar is zero for this spacetime. The component of
acceleration is
\begin{equation}
\dot{u}_1=-\frac{\alpha}{r}.
\end{equation}
The only component of rotation is given by
\begin{equation}
\omega_{01}=2\frac{\alpha}{r}e^{\alpha\ln(\frac{r}{r_0})}.
\end{equation}
The shear component turns out to be
\begin{equation}
\sigma_{01}=-\omega_{01}
\end{equation}
while the shear invariant takes the form
\begin{equation}
\sigma=-4\frac{\alpha^2}{r^2}=4{\dot{u}_1}^2.
\end{equation}
The expansion rate becomes
\begin{equation}
\frac{d\Theta}{d\tau}=\frac{\alpha}{r^2}(\alpha+1).
\end{equation}

\subsubsection{Class $D$ admitting eleven homotheties}

All the kinematical properties turn out to be zero for the first
two solutions given by Eqs.(12) and (13). For the third solution,
all the quantities are exactly similar to the quantities given for
the solutions (8) and (9) except the rate of expansion. The rate
of expansion is given by
\begin{equation}
\frac{d\Theta}{d\tau}=\frac{4}{r^2}.
\end{equation}

\subsection{Kinematic Properties in Non-Co-moving Coordinates}

In this section we shall discuss the kinematical properties of the
self-similar solution in non-co-moving coordinates only in the
radial direction.

\subsubsection{Class $A$ admitting four homotheties}

For this class, the acceleration and rotation for all the
solutions turn out to be the same. The two components of
acceleration are
\begin{equation}
\dot{u}_0=\frac{\alpha}{r}e^{\alpha\ln(\frac{r}{r_0})},\quad
\dot{u}_1=-\frac{\alpha}{r}.
\end{equation}
The only rotation component is
\begin{equation}
\omega_{01}=\frac{\alpha}{r}e^{\alpha\ln(\frac{r}{r_0})}.
\end{equation}
The expansion for the first solution turns out to be
\begin{equation}
\Theta=-\frac{(1+\alpha)}{r}.
\end{equation}
The components of shear become
\begin{eqnarray}
\sigma_{00}=4\frac{\alpha}{r}e^{2\alpha\ln(\frac{r}{r_0})},\quad
\sigma_{11}=\frac{2}{3r}(2\alpha-1),\quad
\sigma_{22}=-a^2\frac{7+\alpha}{3r}e^{2\ln(\frac{r}{a})},\nonumber\\
\sigma_{33}=-\frac{(1+\alpha)}{3r},\quad
\sigma_{01}=-\frac{(1+7\alpha)}{3r}e^{2\ln(\frac{r}{a})}.
\end{eqnarray}
The measure of anisotropy is given by
\begin{equation}
\sigma=\frac{1}{9r^2}(113\alpha^2-14\alpha+53).
\end{equation}
The rate of expansion turns out to be
\begin{equation}
\frac{d\Theta}{d\tau}=-\frac{4}{9r^2}(29\alpha^2-2\alpha+14)
-2\frac{\alpha}{r}.
\end{equation}
For the second solution, given by Eq.(6), the non-zero expansion
becomes
\begin{equation}
\Theta=-\frac{(2\beta+\alpha)}{r}.
\end{equation}
The shear components are
\begin{eqnarray}
\sigma_{00}=4\frac{\alpha}{r}e^{2\alpha\ln(\frac{r}{r_0})},\quad
\sigma_{11}=\frac{4}{3r}(\alpha-\beta),\quad
\sigma_{22}=-\frac{(\alpha+8\beta)}{3r}a^2e^{2\beta\ln(\frac{r}{r_0})},\nonumber\\
\sigma_{33}=-\frac{(\alpha+8\beta)}{3r}e^{2\beta\ln(\frac{r}{r_0})},\quad
\sigma_{01}=-\frac{2}{3r}(4\alpha-\beta)e^{\alpha\ln(\frac{r}{r_0})}.
\end{eqnarray}
The shear invariant turns out to be
\begin{equation}
\sigma=\frac{2}{9r^2}(49\alpha^2+16\alpha\beta+70\beta^2).
\end{equation}
The rate of change of expansion becomes
\begin{equation}
\frac{d\Theta}{d\tau}=-\frac{1}{9r^2}(101\alpha^2+9\alpha
r+62\alpha\beta +64\beta^2+18\beta).
\end{equation}
For the third metric, given by Eq.(7), the expansion becomes the
same as for the first solution. The non-vanishing components of
shear are
\begin{eqnarray}
\sigma_{00}=4\frac{\alpha}{r}e^{2\alpha\ln(\frac{r}{r_0})},\quad
\sigma_{11}=\frac{2}{3r}(2\alpha-1),\quad\
\sigma_{22}=-\frac{1}{3r}(\alpha+1)a^2,\nonumber\\
\sigma_{33}=-\frac{1}{3r}(\alpha+7)e^{2\ln(\frac{r}{r_0})},\quad
\sigma_{01}=-\frac{1}{3r}(8\alpha-1)e^{\alpha\ln(\frac{r}{r_0})}.
\end{eqnarray}
The measure of anisotropy is given by
\begin{equation}
\sigma=\frac{1}{9r^2}(98\alpha^2+16\alpha+53)
\end{equation}
and the expansion rate turns out to be
\begin{equation}
\frac{d\Theta}{d\tau}=\frac{5}{9r^2}(19\alpha^2-2\alpha
-10)-\frac{\alpha}{r}.
\end{equation}

\subsubsection{Class $B$ admitting five homotheties}

Now we discuss the kinematical properties for the solutions given
by Eq.(8)-(10) of the class $B$. The expansion scalar remains the
same for all the solutions in this class and is given by
\begin{equation}
\Theta=-\frac{2}{r}.
\end{equation}
The acceleration and rotation components are same for the metrics
given by Eqs.(8) and (9). The non-zero components of acceleration
are
\begin{equation}
\dot{u}_0=\frac{1}{r}e^{\ln(\frac{r}{r_0})},\quad
\dot{u}_1=-\frac{1}{r}.
\end{equation}
The rotation component turns out to be
\begin{equation}
\omega_{01}=\frac{1}{r}e^{\ln(\frac{r}{r_0})}.
\end{equation}
The non-zero components of shear for the metric, given by Eq.(8),
are
\begin{eqnarray}
\sigma_{00}=\frac{4}{r}e^{2\ln(\frac{r}{r_0})},\quad
\sigma_{11}=\frac{2}{3r},\quad
\sigma_{22}=-\frac{8}{3r}a^2e^{2\ln(\frac{r}{a})},\nonumber\\
\sigma_{33}=-\frac{2}{3r}.\quad
\sigma_{01}=-\frac{7}{3r}e^{\ln(\frac{r}{r_0})}.
\end{eqnarray}
The components of shear for the metric, given by Eq.(9), have the
form
\begin{eqnarray}
\sigma_{00}=\frac{4}{r}e^{2\ln(\frac{r}{r_0})},\quad
\sigma_{11}=\frac{2}{3r},\quad
\sigma_{22}=-\frac{2}{3r}a^2,\nonumber\\
\sigma_{33}=-\frac{8}{3r}e^{2\ln(\frac{r}{r_0})},\quad
\sigma_{01}=-\frac{7}{3r}e^{\ln(\frac{r}{r_0})}.
\end{eqnarray}
The shear invariant is same for the metrics, given by Eqs.(8) and
(9), which turns out to be
\begin{equation}
\sigma=\frac{167}{9r^2}.
\end{equation}
The rate of expansion also remains the same for the two solutions
given by Eqs.(8) and (9) and is the following
\begin{equation}
\frac{d\Theta}{d\tau}=-\frac{188}{9r^2}-\frac{1}{r}.
\end{equation}
For the third solution, given by Eq.(10), the acceleration and
rotation components are zero. The non-zero components of shear are
\begin{eqnarray}
\sigma_{11}=-\frac{4}{3r},\quad
\sigma_{22}=-\frac{8}{3r}a^2e^{2\ln(\frac{r}{a})},\quad
\sigma_{33}=-\frac{8}{3r}e^{2\ln(\frac{r}{a})},\quad
\sigma_{01}=\frac{2}{3r}.
\end{eqnarray}
The measure of anisotropy is
\begin{equation}
\sigma=\frac{140}{9r^2}
\end{equation}
and the expansion rate for this metric turns out to be
\begin{equation}
\frac{d\Theta}{d\tau}=-\frac{152}{9r^2}.
\end{equation}

\subsubsection{Class $C$ admitting seven homotheties}

The solution for this class has two components of acceleration.
These turn out to be
\begin{equation}
\dot{u}_0=\frac{\alpha}{r}e^{\alpha\ln(\frac{r}{r_0})},\quad
\dot{u}_1=-\frac{\alpha}{r}.
\end{equation}
The expansion is given by
\begin{equation}
\Theta=-3\frac{\alpha}{r}=3\dot{u}_1
\end{equation}
and the rotation component is
\begin{equation}
\omega_{01}=\frac{\alpha}{r}e^{\alpha\ln(\frac{r}{r_0})}.
\end{equation}
The components of shear are
\begin{eqnarray}
\sigma_{00}=4\frac{\alpha}{r}e^{2\alpha\ln(\frac{r}{r_0})},\quad
\sigma_{22}=-3\frac{\alpha}{r}a^2e^{2\alpha\ln(\frac{r}{r_0})},\quad
\sigma_{33}=\frac{1}{a^2}\sigma_{22},\nonumber\\
\sigma_{01}=-2\omega_{01}.
\end{eqnarray}
The shear invariant becomes
\begin{equation}
\sigma=30\frac{\alpha^2}{r^2}=30{\dot{u}_1}^2
\end{equation}
and the rate of expansion turns out to be
\begin{equation}
\frac{d\Theta}{d\tau}=-\frac{\alpha}{r^2}(33\alpha+2)-\frac{\alpha}{r}.
\end{equation}

\subsubsection{Class $D$ admitting eleven homotheties}

For the first metric in this class, given by Eq.(12), the
components of acceleration and rotation turn out to be zero. The
expansion becomes
\begin{equation}
\Theta=-\frac{1}{r}.
\end{equation}
The components of shear take the form
\begin{eqnarray}
\sigma_{11}=-\frac{2}{3r},\quad \sigma_{22}=-\frac{a^2}{3r},\quad
\sigma_{33}=-\frac{7}{3r}e^{2\ln(\frac{r}{r_0})},\quad
\sigma_{01}=\frac{1}{3r}.
\end{eqnarray}
The measure of anisotropy is given by
\begin{equation}
\sigma=\frac{53}{9r^2}
\end{equation}
while the expansion rate turns out to be
\begin{equation}
\frac{d\Theta}{d\tau}=-\frac{56}{9r^2}.
\end{equation}
For the second solution, given by Eq.(13), all the quantities
vanish which is exactly similar to the co-moving case.

For the third metric, given by Eq.(14), two non-zero components of
acceleration turn out to be
\begin{equation}
\dot{u}_0=\frac{1}{r}e^{\ln(\frac{r}{r_0})},\quad
\dot{u}_1=-\frac{1}{r}.
\end{equation}
The expansion becomes
\begin{equation}
\Theta=-\frac{1}{r}=\dot{u}_1
\end{equation}
and the rotation component is
\begin{equation}
\omega_{01}=\frac{1}{r}e^{\ln(\frac{r}{r_0})}.
\end{equation}
The shear components turn out to be
\begin{eqnarray}
\sigma_{00}=\frac{4}{r}e^{2\ln(\frac{r}{r_0})},\quad
\sigma_{11}=\frac{4}{3r},\quad
\sigma_{22}=-\frac{1}{3r}a^2,\nonumber\\
\sigma_{33}=-\frac{1}{3r},\quad
\sigma_{01}=-\frac{8}{3r}e^{\ln(\frac{r}{r_0})}.
\end{eqnarray}
The measure of anisotropy is given by
\begin{equation}
\sigma=\frac{98}{9r^2}
\end{equation}
and the rate of expansion yields
\begin{equation}
\frac{d\Theta}{d\tau}=-\frac{101}{9r^2}-\frac{1}{r}.
\end{equation}

\section{Singularities}

In this section, we shall explore the singularities of the
self-similar solutions given by Eqs.(5)-(14). We shall use the
Kretschmann scalar to find the singularities of these solutions.
The Kretschmann scalar is defined by
\begin{equation}
K=R_{abcd}R^{abcd},
\end{equation}
where $R_{abcd}$ is the Riemann tensor. For the solution, given by
Eq.(5), it reduces to
\begin{equation}
K=2\frac{\alpha^2}{r^4}(\alpha^2-2\alpha+2).
\end{equation}
Since $K$ diverges at $r=0$ hence this solution is singular at
$r=0$.

For the second solution, given by Eq.(6), the Kretschmann scalar
becomes
\begin{equation}
K=\frac{2}{r^4}(2\alpha^2\beta^2+3\beta^4+\alpha^4
+\alpha^2(1-2\alpha)+2\beta^2(1-2\beta)).
\end{equation}
This also shows that the solution is singular at $r=0$.

For the third solution, given by Eq.(7), the Kretschmann scalar
turns out to be
\begin{equation}
K=\frac{2}{r^4}\alpha^2(\alpha^2-2\alpha+2)
\end{equation}
which again gives the singularity at $r=0$.

For all the solutions of the class $B$, given by Eqs.(8)-(10), the
Kretschmann scalar remains the same and is given by
\begin{equation}
K=\frac{2}{r^4}.
\end{equation}
It is clear that $K$ diverges at the point $r=0$. Thus the
solutions are singular at $r=0$.

For the solution in class $C$, given by Eq.(11), the Kretschmann
scalar reduces to
\begin{equation}
K=48\frac{\alpha^2}{r^4}(1+2\alpha^3-2\alpha).
\end{equation}
It is obvious from here that $K$ diverges at $r=0$ and
consequently the solution is singular at $r=0$.

For the last class, the Kretschmann scalar turns out to be zero
for all the solutions.

\section{Conclusion}

Self-similar solutions in GR are very important and it is
interesting to discuss their physical features. Keeping this point
in mind, we have explored some kinematic properties and the
singularity feature of such solutions representing cylindrically
symmetric spacetime. We have discussed their properties both in
co-moving as well as in non-co-moving coordinates (only in radial
direction). We have explored acceleration, expansion, rotation,
shear, rate of change of expansion and finally the singularity.

We obtain zero expansion in co-moving coordinates for all the
solutions while in non-co-moving coordinates we have expansion
negative/positive or zero depending upon the value of $\alpha$ and
$\beta$. There is an exceptional solution in the last class, given
by Eq.(13), for which all the kinematical properties become zero
in co-moving coordinates as well as in non-co-moving coordinates.
In co-moving coordinates, it follows only one component of
acceleration whereas there exist two components of acceleration in
non-co-moving coordinates. The rotation component remains the same
in both coordinates except the factor of half in non-co-moving
coordinates. We see that only one component of shear exists in
co-moving coordinates which is equal to negative of rotation in
co-moving coordinates while four or five components of shear exist
in non-co-moving coordinates. The measure of anisotropy is given
by shear scalar and this scalar is much larger in non-co-moving
coordinates than in co-moving coordinates. The rate of change of
expansion is zero for the solutions given by Eqs.(10)-(13). Thus
we can say that these solutions are in the state of equilibrium in
co-moving coordinates. All the non-zero properties of each
solution become infinite at $r=0$.

We find that all the self-similar solutions of the first kind in
classes $A,~B,~C$ are singular at $r=0$ while in class $D$, the
Kretschmann scalar becomes zero. We conclude that when we use
non-co-moving coordinates, the quantities may become more
complicated. However, we get simpler results in co-moving
coordinates.

\begin{description}
\item  {\bf Acknowledgment}
\end{description}

The authors acknowledge the enabling role of the Higher Education
Commission Islamabad, Pakistan and appreciate its financial
support through {\it Merit Scholarship Scheme for Ph.D. Studies in
Science and Technology (200 Scholarships)}.

\vspace{2cm}

{\bf \large References}

\begin{description}

\item{[1]} Penston, M.V.: Mon. Not. R. Astr. Soc.
{\bf144}(1969)425;\\
           Larson, R.B.: Mon. Not. R. Astr. Soc.
           {\bf145}(1969)271;\\
           Shu, F.H.: Astrophys. J. {\bf214}(1977)488;\\
           Hunter, C.: Astrophys. J. {\bf218}(1977)834.

\item{[2]} Cahill, M.E. and Taub, A.H.:   Commun. Math. Phys.
           {\bf21}(1971)1.

\item{[3]} Carter, B. and Henriksen, R.N.:  Annales De Physique
           {\bf14}(1989)47.

\item{[4]} Carter, B. and Henriksen, R.N.:  J. Math. Phys.
           {\bf32}(1991)2580.

\item{[5]} Carr, B.J.: Phys. Rev. {\bf D62}(2000)044022.

\item{[6]} Carr, B.J. and Coley, A.A.: Phys. Rev. {\bf D62}(2000)044023.

\item{[7]} Coley, A.A.: Class. Quant. Grav. {\bf14}(1997)87.

\item{[8]} McIntosh, C.B.G.: Gen. Relat. Gravit. {\bf7}(1975)199.

\item{[9]} Benoit, P.M. and Coley, A.A.: Class. Quant. Grav.
           {\bf15}(1998)2397.

\item{[10]} Carr, B.J., Coley, A.A., Golaith, M., Nilsson, U.S. and Uggla, C.:
           Class. Quant. Grav. {\bf18}(2001)303-324.

\item{[11]} Carr, B.J., Coley, A.A., Golaith, M., Nilsson, U.S. and Uggla, C.:
            Phys. Rev. {\bf D61}(2000)081502.

\item{[12]} Coley, A.A. and Golaith, M.: Class. Quant. Grav. {\bf17}(2000)2557-2588.

\item{[13]} Maeda, H., Harada, T., Iguchi, H. and Okuyama, N.: Phys. Rev.
           {\bf D66}(2002)027501.

\item{[14]} Maeda, H., Harada, T., Iguchi, H. and Okuyama, N.: Prog. Theor. Phys.
           {\bf108}(2002)819.

\item{[15]} Maeda, H., Harada, T., Iguchi, H. and Okuyama, N.: Prog. Theor. Phys.
           {\bf110}(2003)25.

\item{[16]} Qadir, A., Sharif, M. and Ziad, M.: Class. Quant. Grav. {\bf17}(2000)345-349.

\item{[17]} Sharif, M. and Aziz, Sehar: Int. J. Mod. Phys. {\bf D}(2005)\\
           (arxiv:gr-qc/0406029).

\item{[18]} Stephani, H., Kramer, D., Maccallum, M., Hoenselaers, C. and Herlt,E.
            \textit{Exact Solutions of Einstein's Field Equations}
            (Cambridge University Press, 2003).

\item{[19]} Wald, R.M. \textit{General Relativity} (University of Chicago,
            Chicago, 1984).

\end{description}

\end{document}